\documentclass[fleqn,twoside]{article}
\usepackage{espcrc2}
\usepackage{epsfig}

\usepackage[figuresright]{rotating}

\newcommand{\bce}{\begin{center}}
\newcommand{\ece}{\end{center}}
\newcommand{\be}{\begin{equation}}
\newcommand{\ee}{\end{equation}}
\newcommand{\bea}{\begin{eqnarray}}

\newcommand{\eea}{\end{eqnarray}}

\newcommand{\ba}{\begin{array}}
\newcommand{\ea}{\end{array}}

\newcommand{\brho}{\mbox{\boldmath $\rho$}}

\newcommand{\bDelta}{\mbox{\boldmath $\Delta$}}

\newcommand{\bkappa}{\mbox{\boldmath $\kappa$}}

\newcommand{\bb}{{\bf b}}
\newcommand{\br}{{\bf r}}

\newcommand{\bk}{{\bf k}}

\newcommand{\bp}{{\bf p}}
\newcommand{\bq}{{\bf q}}

\def\lsim{\mathrel{\rlap{\lower4pt\hbox{\hskip1pt$\sim$}}
    \raise1pt\hbox{$<$}}}         
\def\gsim{\mathrel{\rlap{\lower4pt\hbox{\hskip1pt$\sim$}}
    \raise1pt\hbox{$>$}}}         
\def\beq{\begin{equation}}
\def\eeq{\end{equation}}
\def\bea{\begin{eqnarray}}
\def\eea{\end{eqnarray}}

\def\lsim{\mathrel{\rlap{\lower4pt\hbox{\hskip1pt$\sim$}}
    \raise1pt\hbox{$<$}}}         

\def\gsim{\mathrel{\rlap{\lower4pt\hbox{\hskip1pt$\sim$}}
    \raise1pt\hbox{$>$}}}         

\newcommand{\AmS}{{\protect\the\textfont2
  A\kern-.1667em\lower.5ex\hbox{M}\kern-.125emS}}

\title{Fate of $k_{\perp}$-factorization for hard processes in nuclear
environment}

\author{N.N. Nikolaev, 
{Institut f. Kernphysik, Forschungszentrum
J\"ulich, 
52425 J\"ulich, Germany\\
and
{L.D.Landau Institute for Theoretical Physics, 
142432 Chernogolovka, Russia}
}}
       
\begin{document}

\begin{abstract} Large thickness of heavy nuclei brings in a new
scale into the pQCD description of hard processes in nuclear environment.
The familiar linear $k_{\perp}$-factorization breaks down and must
be replaced by a new concept of the nonlinear $k_{\perp}$-factorization
introduced in \cite{Nonlinear}. I demonstrate the salient features
of nonlinear $k_{\perp}$-factorization on an example of hard dijet
production in DIS off heavy nuclei. I also discuss briefly the 
non-linear BFKL evolution for gluon density of nuclei.
\vspace{1pc}
\end{abstract}

\maketitle

\section{INTRODUCTION}
The linear $k_{\perp}$-factorization is part and parcel of
the pQCD description of high energy hard processes off free nucleons.
A large thickness 
of a target nucleus introduces a new scale - the so-called saturation 
scale $Q_A^2$, - which breaks the linear $k_{\perp}$-factorization 
theorems for hard scattering in nuclear environment. This property 
can be linked to the
unitarity constraints for the colour dipole-nucleus interaction. 
In this talk I review the recent work by the ITEP-J\"ulich-Landau
collaboration in which a new concept of the nonlinear 
$k_{\perp}$-factorization has been introduced 
\cite{Nonlinear,PionDijet,SingleJet,pAdijet}
and illustrate the major results on an example of dijet production 
in DIS off heavy nuclei.

\section{$k_{\perp}$-FACTORIZATION FOR DIS OFF FREE NUCLEONS}
The parton-fusion approach to shadowing introduced in 1975 \cite{NZfusion} is 
equivalent to the unitarization on the colour dipole-nucleus interaction 
\cite{NZ91}.
One starts with the colour-dipole 
factorization for DIS at small $x \lsim x_A=1/R_A m_N$, when 
the coherency over the thickness of
the nucleus holds for the $q\bar{q}$ Fock states of the virtual photon: 

\bea
&&\sigma_T(x,Q^2) = 
\langle {{\gamma^*}}|
{{\sigma(x,\br)}} |{{\gamma^*}} 
\rangle \nonumber\\
&& =\int_{0}^1 dz \int {{d^2\br}}
{{\Psi^*_{{{\gamma^*}}}(z,{{\br}})}}
 {{\sigma(x,\br)}} {
{\Psi_{{{\gamma^*}}}(z,{{\br}})}}\,.
\nonumber 
\eea
Here $z$ and
$(1-z) $ is the energy partition between {{$q$}}
 \& ${{\bar{q}}}$ and ${{\br}}=$ size of the colour dipole.
There is an  {{equivalence}} between
{{colour dipole}} and
{{$k_{\perp}$-factorization}} \cite{NZ91,NZglue,NZ94}:
\bea
{{\sigma(x,\br)}}&=& \alpha_S(r)
\int {d^2{{\bkappa}}4\pi
[1-e^{i{{\bkappa}} \br }] \over
N_c{{\kappa}}^4} \cdot {{{\partial
G_N \over \partial\log{{\kappa}}^2} }} \,,\nonumber\\
 f(x,{{\bkappa}} ) &=& {4\pi \over
N_c\sigma_0(x)}\cdot {1 \over {{\kappa}}^4} \, \cdot
{{{\partial G_N(x,{{\bkappa}})
\over
\partial\log{{\kappa}}^2} }}\, .\nonumber
\eea
The
$x$-dependence of ${{\sigma(x,\br)}}$ is governed by the 
colour dipole BFKL equation \cite{NZZJETPLett}.
The unintegrated gluon density 
${{f(x,\bkappa)}}$ furnishes a {{universal}} description of 
$F_{2p}(x,Q^2)$ and of the final states. For instance,
the linear {{$k_{\perp}$-factorization}} for forward dijet
cross section reads
\bea
{d\sigma_N \over dz {{d^2\bp_+}}
{{ d^2\bDelta}}} = 
\cdot { \alpha_S({{\bp^2}}) \over 2(2\pi)^2}
 f(x,{{ \bDelta}} )\nonumber\\
\times 
\left|\Psi(z,\bp_+) -
\Psi(z,{{\bp_+}} -
{{\bDelta }})\right|^2 \,,
\label{eq:2.2}
\eea
where  
${{\bDelta}}={{\bp_+}} +
{{\bp_-}}$is the jet-jet decorrelation momentum. 

\section{COLLECTIVE NUCLEAR GLUE}

The colour dipole-nucleus cross-section 
\cite{NZ91} 
$$
{{\sigma_A(\br)}} = 2\!\!\int\!\!
d^2{{\bb} }[1 -\exp(-{1\over 2}
{{\sigma(\br)}}
T({{\bb}}))]$$
defines the collective nuclear glue per
{{ unit area}} in the impact parameter space,
${{\phi}}({{\bb}},{{\bkappa}}
) $
\cite{NSSdijet,Nonlinear}:

\begin{figure}[!t]
\epsfig{file=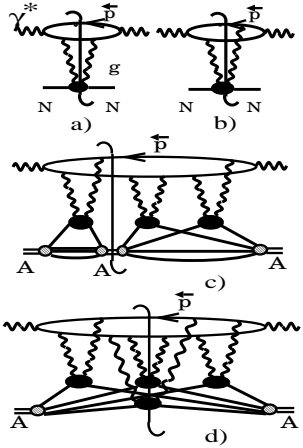,height=6cm, width=4cm}
\caption{ The typical unitarity cuts and dijet final states in DIS : (a),(b) - 
free-nucleon target, (c) - coherent diffractive DIS off a nucleus,
(d) - truly inelastic DIS with multiple colour excitation of the nucleus.}
\label{fig:UnitarityDiffraction2004}
\end{figure}
\bea \Gamma_A({{\bb}},{{\br}})&=&[1 -\exp(-{1\over 2}
{{\sigma(\br)}}T({{\bb}}))]\nonumber\\
&=&\int d^2{{\bkappa}}
{{\phi}}({{\bb}},{{\bkappa}}
) \{1-\exp[i {{\bkappa}}
{{\br}} ]\} \,. \label{eq:2.3} \eea
A useful expansion is
\bea
{{\phi}}({{\bb}},{{\bkappa}}
) &=& \sum_{j=1}^{\infty}{{
w_{j}({{\bb}})}}
 f^{(j)}({{\bkappa}} )\, ,
\nonumber \\
{{w_{j}({{\bb}})}} &=&{
1 \over j!} \left[{1\over 2}T({{\bb}})\right]^j \exp\left[-\nu_{A}({{\bb}})\right],
\nonumber \eea
where $\nu_{A}({{\bb}}) =  
{1\over 2}\sigma_0(x) {{T(\bb)}}$ with $ \sigma_0(x) = {{\sigma(x,\br)}}\big|_{{{r\to \infty}}}$ and
$T(\bb)$ being the optical the
thickness of a nucleus measures the opacity of a nucleus to large dipoles,
 ${{w_{j}}}$ is the probability
to find {{ $j$ overlapping nucleons}} at impact
parameter ${{\bb}}$ in a Lorentz-contracted
nucleus and  $f^{(j)}$is a {{collective glue of $j$
overlapping nucleons}}:
\bea f^{(j)}({{\bkappa}} )\,=\,  \int
\prod_{i}^j d^2{{\bkappa}} _{i}
f({{\bkappa}} _{i})
\delta({{\bkappa}} -\sum_{i}^j
{{\bkappa}} _i).
\nonumber \eea
The plateau at small momenta of gluons, 
\bea
{{\phi}}({{\bb}},{{\bkappa}})
&\approx& {1\over \pi}{{{ {{
Q_A^2}}}}({{\bb}}) \over
({{\bkappa}}^2+ {{
Q_A^2}}({{\bb}}))^2}\, , \nonumber\\
{{ Q_A^2}}({{\bb}},x) &\approx&
 {
4\pi^2 \over N_c} \alpha_S({{
Q_A^2}})G(x,{{ Q_A^2}}) T({{\bb}})\,,
\label{eq:C.12} \eea
is a signal of the saturation effect.
The collective nuclear glue furnishes the linear
$k_{\perp}$-factorization representation  
for DIS off nuclei,
\bea
\sigma_{{{\gamma^*}}A}& =& 
\int {{d^2\bb}}\langle {{\gamma^*}} | 2\{1- \exp[-{1\over 2}{{\sigma({{\br}})}}{{T(\bb)}}]\}| 
{{\gamma^*}}\rangle 
\nonumber\\
&=&\int {{d^2\bb}} \int {d^2{{\bp}} \over (2\pi)^2} \alpha_S({{\bp}}^2)\nonumber\\
&\times&
\int  d^2{{\bkappa}}  {{{{\phi}}}}({{\bkappa}} )
(\Psi(z,{{\bp}}) - \Psi(z,{{\bp}}-{{\bkappa}} ))^2
\label{eq:8.3}
\eea
which is exactly the same as for the nucleon target, subject to
$f({{\bkappa}}) {{\Longleftrightarrow}}
 {{{{\phi}}}}({{\bkappa}})$.

\section{NON-ABELIAN INTRANUCLEAR EVOLUTION OF COLOUR DIPOLES}
 
The two typical final states in DIS off heavy nucleus are shown
in fig.~\ref{fig:UnitarityDiffraction2004}. The coherent diffraction
with large rapidity gap between the target nucleus in the ground state
and diffractive hadronic 
debris of the photon makes $\approx 50\%$ of the total cross section
\cite{NZZdiffr}
and gives exactly back-to-back correlated dijets. In the 
truly inelastic DIS with multiple colour excitation of the nucleus
one encounters the non-Abelian intranuclear evolution of colour 
dipoles, the consistent description of which 
based on the ideas from 
\cite{SlavaPositronium,NPZcharm} is found in \cite{Nonlinear} .
\noindent\begin{figure}[!t]
   \centering
   \epsfig{file=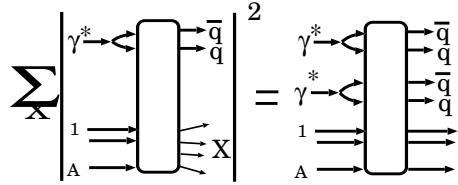,width=6cm,height=2.5cm}
\caption{Unitarity diagram for the dijet spectrum in terms
of the 4-parton scattering amplitude.}
\label{fig:4BodyUnitarity}
\end{figure}
\noindent
Specifically, the ab 
initio calculation of the nuclear distortion of the
two-parton density matrix the Fourier transform of which gives the
spectrum of dijets, can be reduced, upon the closure over nuclear excitations, 
to the problem of intranuclear
propagation of the colour-singlet 4-parton states as illustrated
in fig. \ref{fig:4BodyUnitarity}: 

\bea
{d\sigma_{in} \over dz d^2{{\bp_+}} d^2{{\bp_-}}} =
{1\over (2\pi)^4} \int d^2 {{\bb_+}}' d^2{{\bb_-}}' 
d^2{{\bb_+}} d^2{{\bb_-}} \nonumber\\
\times \exp[-i{{\bp_+}}({{\bb_+}} -{{\bb_+}}')-i{{\bp_-}}({{\bb_-}} -
{{\bb_-}}')]\nonumber\\
\Psi^*(Q^2,z,{{\bb_+}}' -{{\bb_-}}')
\Psi(Q^2,z,{{\bb_+}} -{{\bb_-}})\nonumber\\
\times \Bigl\{S_{4A}({{\bb_+}}',...,{{\bb_-}})- S_{4A}^{(Diffr)}({{\bb_+}}',
...,{{\bb_-}}) \Bigr\}\, ,
\nonumber
\eea
where we subtracted the diffractive contribution. To
the standard dilute-gas nucleus approximation, the
Glauber-Gribov theory gives
 \bea S_{4A}({{\bb_+}}',...,{{\bb_-}})=\exp\{-
{1\over 2}\sigma_{4}({{\bb_+}}',...,{{\bb_-}})
T({{\bb}})\}\, .
\nonumber\label{eq:3.9} 
\eea 
where $\sigma_{4}$ is the coupled-channel operator
in the space of {{singlet-singlet}} $|{{11}}\rangle$
 or  {{ octet-octet}} $|{{88}}\rangle$
4-body dipoles, see
ref. \cite{Nonlinear} for more details.

\section{THE FATE OF $k_{\perp}$-FACTORIZATION FOR NUCLEAR TARGETS}
The single-quark spectrum in DIS exhibits the Abelianization  property.
Namely, the truly inelastic, 
\bea
{d \sigma_{in}\over d^2\bb d^2{{\bp}} dz }   =  
{1 \over (2\pi)^2}~~~~~~~~~~~
~~~~\nonumber\\
\times
\Bigl\{
\int  d^2{{\bkappa}} 
{{\phi({{\bkappa}} )}}
\left|\Psi(z,{{\bp}}) -
\Psi(z,{{\bp}}-
{{\bkappa}}) \right|^2  \nonumber\\
-
 {{\Bigl|}}
\int d^2{{\bkappa}}
{{\phi({{\bkappa}})}} 
(\Psi(z,{{\bp}}) -
\Psi(z,{{\bp}}-
{{\bkappa}})){{\Bigr|^2}}\Bigr\}
\label{eq:Inel}
\eea
and coherent diffractive 
\bea
{d \sigma_{D}\over d^2\bb d^2{{\bp}} dz }   =  
{1 \over (2\pi)^2}
~~~~~~~~~~~~~~~~~~\nonumber\\
\times 
\left|\int d^2{{\bkappa}}{
{\phi({{\bkappa}})}} 
(\Psi(z,{{\bp}}) -
\Psi(z,{{\bp}}-{{\bkappa}}))\right|^2 \, .
\label{eq:DifDIS}
\eea
spectra add to precisely eq.~
\ref{eq:8.3}. I.e., the linear $k_{\perp}$-factorization in terms of 
the collective  nuclear glue holds, which is a
feature of DIS where the photon is a colour
singlet projectile. The same is not true  
for other projectiles,
see Sch\"afer's talk at this conference 
\cite{WolfgangDiffraction2004}.

The nuclear dijet spectrum takes a simple form in the large-$N_c$ approximation:
\bea
{d\sigma_{in} \over d^2{{\bb}} dz d\bp_{-} d{{\bDelta}}} = {1\over
2(2\pi)^2} \alpha_S \sigma_0 T({{\bb}})
\int_0^1 d {{\beta}} \nonumber\\
\int d^2{{\bkappa_3}} d^2{{\bkappa}} f({{\bkappa}})
{{\Phi}}({{\beta}}\nu_A({{\bb}}),{{\bDelta}} -{{\bkappa_3}} -{{\bkappa}})
\\
\label{eq:Dijet1} 
{{\Phi}}({{\beta}}\nu_A({{\bb}}),{{\bkappa_3}})
{{\Bigl|}}
\int d^2{{\bkappa_1}}{{\Phi}}({{(1-\beta)}}\nu_A({{\bb}}),{{\bkappa_1}})\nonumber\\
\bigl\{\Psi(z,\bp_{-} +{{\bkappa_1}} +{{\bkappa_3}}) -\nonumber\\
- 
\Psi( z,\bp_{-} +{{\bkappa_1}} +{{\bkappa_3}}+{{\bkappa}})\bigr\}
{{\Bigr|^2}}
\, .\nonumber
\eea
It is a  manifestly {{nonlinear}} functional of the
collective nuclear glue and here emerges a concept of the
nonlinear $k_{\perp}$-
{{factorization}}.
The slice ${{(1-\beta)}}$ in which
the dipole was in the colour-singlet
state  gives the  
{{Initial State Interaction (distortion of the WF)}}
The slice 
{{
${{\beta}}$}} in which
the dipole is in the colour-octet state 
gives the Final State Interaction. In DIS it looks as an 
independent broadening of the quark
and antiquark jets, such a symmetric form no longer holds
for pQCD subprocesses of relevance to $pA$ collisions as 
the interplay of ISI and FSI becomes more involved; still 
the generic structure of the spectra is similar to (\ref{eq:Dijet1})
\cite{pAdijet}. 
\begin{figure}[!t]
\begin{center}
   \epsfig{file=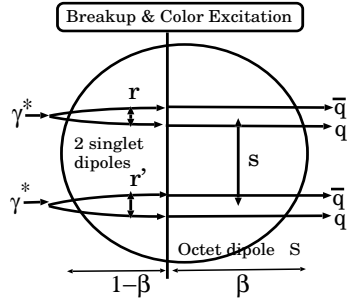,width=4.5cm, height=4cm}
\caption{The colour excitation of the dipole in the
large-$N_c$ approximation.}
\end{center}
\end{figure}

For hard dijets, $|\bp_\pm|^2 \gsim Q_A^2$, the dijet spectrum take 
the convolution form:
 
\bea {d\sigma_{in} \over d^2{{\bb}} dz
d^2{{\bp_+}} d^2{{\bDelta}}}=
T({{\bb}}) \times \int
d^2{{\bkappa}} \int_0^1 d
{{\beta}}\nonumber\\
\times
{{\Phi}}(2{{\beta}} \lambda_c
\nu_A({{\bb}}),{{\bDelta}} -
{{\bkappa}}) {d\sigma_{N} \over dz
d^2{{\bp_+}} d^2{{\bkappa}} }\, .
\label{eq:Dijet2}
\eea
where  $$
{{\Phi}}(\nu_{A}({{\bb}}),
{{\bkappa}})=
\exp(-\nu_{A}({{\bb}}))\delta^{{{(2)}}}
({{\bkappa}}) + {{\phi}}(\nu_{A}({{\bb}}),{{\bkappa}})$$
and $\lambda_c \equiv
C_A/2C_F $. 
The convolution form (\ref{eq:Dijet2}) makes the nuclear
enhancement of decorrelation obvious one. 
{{Semihard}} dijets,
$|\bp_\pm|^2 \lsim Q_A^2$, are completely decorrelated.

The work on applications to the centrality dependence of single-jet
spectra to dijet decorrelations at RHIC \cite{STAR,BRAHMS} is in progress.
  
\section{SMALL-x EVOLUTION OF COLLECTIVE NUCLEAR GLUE.}

Despite the manifest breaking of the linear
$k_{\perp}$-factorization,
the collective nuclear glue remains a useful concept. 
For a free nucleon
target 
 the effect of the $q\bar{q}g$ and higher Fock states in the photon
{{is reabsorbed}} in 
the {{linear}} BFKL evolution for 
the dipole cross section with the photon 
treated as the $q\bar{q}$ state.
One possible definition of the nonlinear 
BFKL evolution for nuclear glue is to
insist on the same representation for nuclear
cross section. It is indeed possible although
without a closed-form 
evolution equation. 

We comment here on  
{{first correction}} 
 $\propto \log{1\over x}$ to the nuclear 
profile function and nuclear collective glue. The correction 
 $\delta \Gamma_A(x,{{\bb} },{\br})$
to the colour dipole-nucleus profile function for the
$q\bar{q}g$ Fock state in the photon equals
\bea
\int d^2{{\bb} } 
{\partial \delta \Gamma_A(x,{{\bb} },{\br}) 
\over \partial \log{1\over x}} =  
K_0 \int d^2{{\brho}} 
{{{\br}}^2 \over {{\brho}}^2
({{\brho}}-{{\br}})^2}
\nonumber\\\times 2\int d^2{{\bb} } 
[\Gamma_{3A}({{\bb}},{{\brho}},{{\br}}) 
-\Gamma_{A}({{\bb}},{{\br}})]
\eea
\bea
 \Gamma_{3A}({{\bb} },{{\brho}},
{{\br}})&=& 1- 
S_{3A}({{\bb} },{{\brho}},
{{\br}}) \nonumber\\
&=& 
1-\exp[-{1\over 2}\sigma_3({{\brho}},{{\br}})
T({{\bb} })]
\nonumber
\eea
where $\sigma_3({{\brho}},{{\br}})$ is the 3-parton cross
section \cite{NZ94}.

A simplified Glauber-Gribov formula holds at large-$N_c$,
$
S_{3A}({{\bb} },{{\brho}},
{{\br}}) =S_{2A}({{\bb} },
{{\brho}}-{{\br}})
S_{2A}({{\bb} },{{\brho}}).
$
Here  ${\partial \delta \Gamma_A(x,{{\bb} },{\br})/
\partial \log{1\over x}}$ is a {{nonlinear}} 
functional of $\Gamma_{2A}$, the 
identification of $\Gamma_A(x,{{\bb} },{{\br}})$
with  $\Gamma_{2A}(x,{{\bb} },{\br})$, and
the 
extension of the {{first
iteration}} to what has become known as the closed-form 
Balitsky-Kovchegov nonlinear equation  
\cite{BalitskiKovchegov} is  
{{unwarranted}}.
In terms of the nuclear transparency for large dipoles, $
S_{A}({{\bb} },\sigma_{0})=\exp[-{1\over 2}
\sigma_0 T({{\bb} })], $ 
the {{first correction }} to the  unintegrated nuclear glue
takes the form
\bea
{\partial {\delta\phi_A(x,{{\bb} },\bDelta)} 
\over \partial \log{1\over x}} =S_{A}({{\bb} },\sigma_{0}) 
{{{\cal K}_{BFKL}}}\otimes 
{{\phi({{\bb} },\bDelta)}}\nonumber\\
+K_0\int d^2 \bp d^2\bk{{\phi}}({{\bb} },\bk)\Bigl\{K({{\bDelta}}+\bp,
{{\bDelta}}+\bk){{\phi}}({{\bb} },\bp)
\nonumber\\
-
K(\bp,\bp+{{\bDelta}}+\bk){{\phi({{\bb} },\bDelta)}}
\Bigr\}\nonumber\\
=S_{A}({{\bb} },\sigma_{0}) 
{{{\cal K}_{BFKL}}}\otimes 
{{\phi({{\bb} },\bDelta)}} + \nonumber\\ 
+{{{\cal K}_{NonLin}}}\bigl[ 
{{\phi({{\bb} },\bDelta)}}\bigr]
\label{eq:BFKL1}
\eea
where
$K(\bp,\bk) = (\bp-\bk)^2/\bp^2\bk^2$. It contains an {{absorption 
suppressed linear BFKL  term}} with the familiar
kernel  ${{{\cal K}_{BFKL}}}$ \cite{BFKL}. 
For central DIS off heavy nuclei $S_{A}\to 0$ and
 evolution is entirely
driven by the nonlinear term quadratic in ${{\phi}}(\bk)$.

Making use of an explicit form of $K(\bp,\bk)$, one can recast 
(\ref{eq:BFKL1}) for the leading conformal twist nuclear glue in an alternative 
form 
\bea {\partial \delta \phi_A(x,{{\bb} },{{\bDelta}}) \over \partial \log(1/x)} =
{\cal{K}}_{BFKL}\otimes \phi({{\bb} },{{\bDelta}}) +
{{{\cal{Q}}}}[\phi]({{\bb} },{{\bDelta}}) \, . 
\label{eq:BFKL2}
\eea
Here the linear term evolves with the conventional BFKL kernel,
whereas the nonlinear term takes a particularly simple form
\bea {{{\cal{Q}}}}[\phi]({{\bb} },{{\bDelta}}) = 
 -{2 K_0 \over {{\bDelta}}^2} \Big[ \int_{{{\bDelta^2}}} d^2 \bq
\phi({{\bb} },\bq) \Big]^2 \nonumber\\
- 2K_0 \phi({{\bb} },{{\bDelta}}) \int_{{{\bDelta}}^2} {d^2\bp
\over \bp^2} \int_{\bp^2} d^2\bq \phi({{\bb} },\bq)  .\nonumber
\eea
For hard gluons, {{$\bDelta^2 > Q_A^2$}}, one can use an
approximation
$
\phi(\bq) \sim \phi({{\bDelta}}) \left({{{\bDelta}}^2 / q^2}\right)^2
$
with the result 
\bea {{{\cal{Q}}}}[\phi]({{\bDelta}};\bb) \approx -4 K_0 \cdot {{\bDelta}}^2
\phi^2 ({{\bDelta}})\propto 
{\phi({{\bDelta}})\over {{\bDelta}}^2}\,.
\nonumber
\eea
The nonlinear component in (\ref{eq:BFKL2}) gives a pure higher
twist contribution. It doesn't exhaust the nuclear higher twist
terms, though, because the one is contained also in $\phi(x,{{\bDelta}};\bb)$,
see the discussion in \cite{Nonlinear,NSSdijet}.
The character of nonlinearity in terms of 
$ G_A({{\bb}},x,{{Q}})$ is instructive:
\bea
{\partial^2{{\delta G_A}}({{\bb}},x,{{Q}})
\over
\partial\log(1/x)\partial \log{{Q^2}} } ={\cal{K}}_{BFKL}\otimes 
{\partial{{G_A}}({{\bb}},x,{{Q}})
\over
\partial\log{{Q^2}} }\nonumber\\
-{4\alpha_S({{Q}}^2) T({{\bb}})
\over{{Q}}^2} \cdot \left({\partial{{G_A}}({{\bb}},x,{{Q}})
\over
\partial\log{{Q^2}} }\right)^2
\nonumber
\eea

Now have  a look at the plateau region of soft gluons , ${{\bDelta^2}} \ll Q_A^2$.
Here eq.~(\ref{eq:BFKL1}) takes the form
\bea
{\partial {\delta\phi_A(x,{{\bb} },\bDelta)} 
\over \partial \log{1\over x}} =-2C\pi K_0 
{{\phi(x,{{\bb} },0)}}\label{eq:BFKLsoft}
\eea
where the constant factor, $C \sim 1$, depends on the form of the collective 
nuclear glue. If we recall that
$$
{{\phi(x,{{\bb} },0)}} \sim {1 \over \pi 
{{Q_A^2}}({{\bb} },x)}
$$
then (\ref{eq:BFKLsoft}) entails an
{{expansion of the plateau}} width with the decrease of $x$:
$$
 {{Q_A^2}}({{\bb} }) 
\Longrightarrow  {{Q_A^2}}({{\bb} })
\left[ 1+2C\pi K_{0} \log{1\over x} \right] 
$$
The full-fledged nonlinear evolution will be in effect for 
soft-to-hard gluon momenta ${{\bDelta^2}} \lsim Q_A^2$.

\section{CONCLUSIONS}
Nuclear saturation is a straightforward consequence of 
{{opacity of heavy nuclei}} to large colour dipoles. 
The imposition of unitarity constraints within the colour-dipole
approach leads to a unique definition and expansion of nuclear 
unintegrated glue in terms of {{\emph{the collective
glue of overlapping nucleons}}}. The problem of {{\emph{a non-abelian}}} 
intranuclear evolution of colour dipoles has been solved and 
consistent description of single-jet and dijet production in DIS
off nuclei has been developed. We have proven the 
{{\emph{breaking $k_{\perp}$ factorization}}} and instead 
formulated the {{\emph{nonlinear $k_{\perp}$-factorization}}} for forward dijet
production in DIS. The formalism is readily extendible to proton-nucleus
collision in the kinematic conditions of the RHIC experiments, the
corresponding work is in progress. We applied our technique to
the nonlinear BFKL evolution of collective nuclear glue and 
explored the twist properties of the nonlinear component of this
equation. The work an applications to jet
production in $pA$ collisions at RHIC is in progress. 
\\

\noindent
{\bf Acknowledgements:} I'm grateful to R.~Fiore and B.~L\"ohr for the 
invitation to this exciting meeting.

\end{document}